# Evidence for a quantum dipole liquid state in an organic quasi-two-dimensional material

Nora Hassan, [1]  Streit Cunningham, [1]  Martin Mourigal, [2]  Elena I. Zhilyaeva, [3]
Svetlana A. Torunova, [3]  Rimma N. Lyubovskaya, [3]  John Schlueter, [4,5] Natalia Drichko [1*]

[1]The Institute for Quantum Matter and the Department of Physics and Astronomy,
The Johns Hopkins University,
Baltimore, Maryland 21218, USA

[2]School of Physics, Georgia Institute of Technology,
Atlanta, GA 30332, USA

[3]Institute of Problems of Chemical Physics,
Chernogolovka, Russia

[4] Division of Materials Research, National Science Foundation,
Alexandria, VA 22314, USA

[5] Materials Science Division, Argonne National Laboratory, Argonne, IL 60439, USA

[*]To whom correspondence should be addressed; E-mail: drichko@jhu.edu.

**Mott insulators are commonly pictured with electrons localized on lattice sites. Their low-energy degrees of freedom involve spins only. Here we observe emerging charge degrees of freedom in a molecule-based Mott insulator $\kappa$-(BEDT-TTF)$_2$Hg(SCN)$_2$Br, resulting in a quantum dipole liquid state. Electrons localized on molecular dimer lattice sites form electric dipoles that do not order at low temperatures and fluctuate with frequency detected experimentally in our Raman spectroscopy experiments. The heat capacity and Raman scattering response are consistent with a scenario where the composite spin and electric dipole degrees of freedom remain fluctuating down to the lowest measured temperatures.**

Fluctuating dipolar degrees of freedom have been predicted to appear in molecular-based Mott insulators [1, 2] and optical lattices of dipolar molecules [3], and to lead to a spin liquid state in the presence of charge-spin coupling. In quantum paraelectrics fluctuations of electric dipoles are observed in the vicinity of a ferroelectric transition [4]. A quantum dipole liquid in an antiferroelectric on a triangular lattice was recently reported for $BaFe_{12}O_{19}$ [5], but as a band insulator this compound is non-magnetic. A quantum dipole liquid in a Mott insulator is a paradigm for quantum states of matter that brings together quantum paraelectrics and spin liquids. Here we discuss an experimental evidence for the quantum dipole liquid state in a layered organic Mott insulator $\kappa$-(BEDT-TTF)$_2$Hg(SCN)$_2$Br (here BEDT-TTF stands for a molecule bis(ethylenedithio)tetrathiafulvalene). In a presence of charge-spin coupling it may result in a spin liquid state.

Electronic and magnetic phenomena observed in this class of materials are determined by the properties of the molecular-based cation layers (Fig 1A). In a Mott insulator of this origin, electrons are localized on the lattice sites of dimers (BEDT-TTF)$_2^{1+}$ with S=1/2 per site. These dimer sites form layers (Fig. 1B) that can be represented by a two-dimensional anisotropic triangular lattice (Fig. 1C). In most such compounds the dimers have an inversion center and thus zero electric dipole moment. Frustration of the lattice, competing electronic correlations, and magnetic interactions can lead to charge being distributed non-symmetrically between the two molecules in a dimer [6, 1], producing a dipole. This can lead to a broken symmetry ground state, so called "paired electron crystal" [6] or "quantum dipole solid" [1] where the dipoles order forming a ferroelectric state (Fig. 1D) [6]. In contrast to a displacive ferroelectric, a change in the underlying lattice is not necessary in this case. It was suggested [1] that the dipoles can fluctuate in a quantum dipole liquid (Fig. 1E) providing an explanation of the spin liquid state observed in a triangular lattice $\kappa$-(BEDT-TTF)$_2$Cu$_2$(CN)$_3$. However, the evidence for fluctuating quantum dipoles in this material remains elusive [7, 8].

Here we elucidate the properties of the quantum dipole liquid state in the triangular lattice Mott insulator [9, 10] $\kappa$-(BEDT-TTF)$_2$Hg(SCN)$_2$Br ($\kappa$-Hg-Br) (T$_{MI}$= 80 K [11, 12]) by comparing them to those of an isostructural compound $\kappa$-(BEDT-TTF)$_2$Hg(SCN)$_2$Cl ($\kappa$-Hg-Cl), which shows signatures of a quantum dipole solid below 30 K [11, 10]. We first use Raman molecular vibrational spectroscopy to follow the evolution of the distribution of charge on the lattice of these systems through the metal-insulator transition. The on-molecule charge is probed by measuring the frequency of the central C=C molecular bond vibration ($\nu_2$) (Fig. 2E), which changes by ~ -140 cm$^{-1}$ when the charge state changes from (BEDT-TTF)$^0$ to (BEDT-TTF)$^{1+}$ [13, 14]. This is a result of a lengthening of the central C=C bond of the molecule when more charge occupies the highest occupied molecular orbital (HOMO). To exclude effects other than a change of charge distribution on the lattice, we compare the temperature dependence of the parameters of the charge-sensitive $\nu_2$ band to that of the $\nu_3$ mode at 1470 cm$^{-1}$. The latter involves the stretch of the off-center C=C bonds (Fig. 2E) and is not sensitive to charge on the molecule, but would be affected by structral disorder or symmetry breaking. $\nu_3$ band stays a narrow single line throughout the measured temperature range for both materials. For $\kappa$-Hg-Cl a single $\nu_2$ band at about 1490 cm$^{-1}$ is observed in the high temperature metallic state, whereas in the insulating state below $T_{CO}$=30 K $\nu_2$ is split into two bands at 1475 and 1507 cm$^{-1}$ (Fig. 2A). The difference in frequencies of the two modes is much higher than expected for a structural phase transition, but corresponds to charges redistributed within the (BEDT-TTF)$_2^{1+}$ dimer as $+0.4e$ on one molecule, and $+0.6e$ on the other [14, 10]; this breaks the inversion symmetry in a (BEDT-TTF)$_2^{1+}$ dimer, creating an electric dipole. This result confirms that the ground state of κ-Hg-Cl is an order of electric dipoles localized on (BEDT-TTF)$_2^{1+}$ dimer sites [10], the so-called dipole solid [1].

In contrast, for κ-Hg-Br a single $\nu_2$ mode observed in the whole studied temperature range (Fig. 1C) suggests a symmetric dimer with both molecules carrying half a hole (BEDT-TTF)$^{0.5+}$ on average. However, the width of the $\nu_2$ band shows abnormal behaviour on cooling. Line widths of phonons are determined by decay mechanism [15], disorder and dynamics of the lattice and charge systems. For example, the width of the $\nu_3$ vibrational band of BEDT-TTF molecule is determined by decay processes into lower-frequency modes and decreases down to about 5 cm$^{-1}$ at 10 K (see Fig. 2D ). In contrast, the width of $\nu_2$ goes through a minimum of 16 cm$^{-1}$ at around 80 K, and increases again up to 20 cm$^{-1}$ at 10 K ( Fig. 2, C and D ). Adnormal temperature dependence of the width observed only for $\nu_2$ rules out the possibility that structural changes or structural disorder are at its origin. Another possible reason for an increased line width is charge fluctuations. [16, 17].

We estimate the effects of charge fluctuations on the shape of the $\nu_2$ vibration using a "two- site jump" model [16, 17] (see Eq. S1). In this model, we consider $+0.6e$ and $+0.4e$ charged molecules observed in the ordered state of k-Hg-Cl as two static species. They are characterized by frequencies of $\nu_2$ vibrations $\nu_2$ [BEDT-TTF $^{0.4+}$] = 1507 cm$^{-1}$ and $\nu_2$[BEDT-TTF $^{0.6+}$]= 1475 cm$^{-1}$ . Their natural width $\Gamma$ depends on temperature through the lifetime of the measured excited state and is expected to be the same as that of $\nu_3$. The system can jump between these two states with a frequency of $\omega_{EX}=1/\tau$, where $\tau$ is the life time of each state defined by the exchange. As the exchange rate $\omega_{EX}$ between these two states increases, the shape of the resulting spectra changes (Fig. 2B). The two original bands get wider and the difference in positions between them decreases, and at high enough rate $\omega_{EX}$ they merge into a single band. The calculated spectrum at $\omega_{EX}= 0$ reproduces the doublet shape of $\nu_2$ for $\kappa$-Hg-Cl in the dipole solid state (Fig. 2, A and B ). The spectra calculated for $\omega_{EX} =$ 40 and 30 cm$^{-1}$ reproduce the shape of the $\nu_2$ band in $\kappa$-Hg-Br spectra, where the width of $\nu_2$ increases from 16 to 20 cm$^{-1}$ on cooling below 80 K and the band gains slight asymmetry (Fig. 2, B and C ). These results suggest that in $\kappa$-Hg-Br charges fluctuate between two molecules in a dimer with frequency $\omega_{EX}$ that slightly decreases on cooling. In other words, electric dipoles in $\kappa$-Hg-Br fluctuate with this frequency, forming a quantum dipole liquid state.

Apart from the difference in the $\nu_2$ band behaviour, the phonon Raman spectra of $\kappa$-Hg-Br and $\kappa$-Hg-Cl are very similar (Fig. 3, A, B, and S1 A, B ), because the two compounds have very similar crystal structures. While the spectral region below about 200 cm$^{-1}$ for k-Hg-Cl shows only phonon bands (Fig. 3 B), in the $A_{1g}$ scattering channel for k-Hg-Br the phonons are superimposed on a much wider feature with a maximum around 40 cm$^{-1}$ (Fig. 3 A). In $B_{1g}$ scattering channel we also observe this feature, with phonon bands showing broad assymetric shapes, apparently as a result of electron-phonon coupling (Fig. 3 A, S2 B). This asymmetric feature with a maximum around 40 cm$^{-1}$ (see spectra with phonons subtracted in Fig. 3C ) with the width at half maximum of about 40 cm$^{-1}$ gains intensity below 100 K, when $\kappa$-Hg-Br enters the insulating state, and shows weak softening at the lowest temperature. Apparently, this wide feature observed only in $\kappa$-Hg-Br spectra originates from a different scattering channel than phonons. Other potential scattering channels are electronic or magnetic excitations on a triangular lattice of (BEDT-TTF)$_2^{1+}$ dimers. On a triangular lattice polarizations of electronic or magnetic excitations cannot be completely disentangled to elucidate the origin of the excitations, in contrast to a square lattice [18].

Magnetic excitations are expected in the Raman response of a Mott insulator with ordered spins or even spins developing short range correlations [18, 19, 20, 21]. In Mott insulators based on (BEDT-TTF)$_2^{1+}$ dimers magnetic excitations are observed both in Raman spectra of an antiferromagnetically ordered state on a square lattice [19] and in a spin liquid candidate on triangular lattice κ-(BEDT-TTF)$_2$Cu$_2$(CN)$_3$ (Fig. 3E). The spectra of the latter show a continuum of magnetic excitations below 600 cm$^{-1}$. The position of the continuum is defined by the value of $J$ and geometry of the lattice [18]. For κ-(BEDT-TTF)$_2$Cu$_2$(CN)$_3$ it is in agreement with Hubbard-model-based calculations for the magnetic response of S=1/2 on an anisotropic triangular lattice with $J_M$=250 K [20].

It is clear at this point that magnetic interactions in a dipole solid, and possibly quantum dipole liquid would be re-normalized in comparison to a simple (BEDT-TTF)$_2^{1+}$ dimer Mott insulator with charge symmetrically distributed on a dimer. Reference [1] proposes a re-normalization and a decrease of $J$ in a quantum dipole liquid compared to a simple dimer Mott insulator, however without estimating $J$ values. A simple argument suggests that, in a dipole solid, magnetic interactions occur between charge-rich molecules of the neighboring dimers, whereas in a simple Mott insulator the interactions occur between dimer lattice sites, (illustrations in Fig. 3E). An estimate provided by a tight-binding approximation as $J = \frac{4t^2}{U}$, where $t$ is a transfer integral, and $U$ is on-molecule Coulomb repulsion, yields the value of about $J_{DS} = 80K$ for a dipole solid. This is considerably smaller than $J_M = 250K$ [22] for a simple dimer Mott insulator, where the on-dimer $U$ defines magnetic interactions. Here the Coulomb repulsion parameters, as well as transfer integrals are estimated from the optical conductivity spectra [23], and the difference is produced mainly by a variation between the values of $U$ in these two models. A lower $J$ would result in a lower ordering temperature, and a spectrum of magnetic excitation shifting to lower frequencies. However, the maximum of the observed background is around 40 cm$^{-1}$ is found below the expectant $J_{DS}$ value, which is too low in frequency to be interpreted as purely magnetic excitations.

Another possibility is assigning this mode as a collective excitation associated with dipole fluctuations. Dipole fluctuations with frequency of about $\omega_{EX}$ = 40 cm$^{-1}$ are detected through the line shape analysis of charge-sensitive vibrations. If these fluctuations are a collective phenomenon, we would expect a collective mode at about 40 cm$^{-1}$. The low-frequency mode observed in $\kappa$-Hg-Br thus is a good candidate for a collective response of dipole fluctuations. Optically detected collective modes associated with charge fluctuations are found in the metallic state close to a charge ordering metal-insulator transition in organic conductors [24, 25] and in under-doped high temperature cuprate superconductors [26]. In an insulating state, the closest analogy would be a soft mode close to the transition into the ferroelectric state in displacive ferroelectrics such as SrTiO$_3$. A comparatively small width of the band, as well as its increase in intensity and its low-frequency shift below T$_{MI}$=80 K distinguishes it from a Boson peak observed in glasses [27] and supports an interpretation in terms of a fluctuating system of dipoles versus charge glass. An absence of glassy behavior is also supported by the low frequency dielectric response of $\kappa$-Hg-Br [12].

Fluctuations of electric dipoles coupled to S=1/2 spins on a triangular lattice of

(BEDT-TTF)$_2^{1+}$ dimers have been suggested as a mechanism for spin-liquid behaviour [1, 2]. Reference [2] discusses the coupling between the dipole and magnetic degrees of freedom within the Kugel-Khomskii model, showing an analogy between the fluctuating dipole liquid and orbital liquid [28, 29]. This model suggests that for certain values of frustration J'/J and spin-charge coupling $K$, spin order is destabilized, and would produce mixed spin-charge excitations. To understand if the collective mode observed in κ-Hg-Br Raman spectra originates purely in dipole fluctuations or in mixed charge-spin excitations, theoretical calculations of the excitation spectrum for such a system would be of great importance.

Our heat capacity data suggest the presence of itinerant excitations in κ-Hg-Br but not in κ-Hg-Cl. The heat capacity $C_p$ of κ-Hg-Br and κ-Hg-Cl was measured in the temperature range between 40 K and 100 $m$K. The temperature dependences of heat capacity for these compounds overlap within the error of the measurements in the temperature range above 6 K (Fig. 3D), excluding the feature at 30 K in the κ-Hg-Cl data indicating a charge order transition. The low temperature heat capacity $C_p = \beta T^3 + \gamma T$ of both compounds shows basically the same bosonic contribution $\beta$ = 19.0 ± 2.5 mJ K$^{-4}$ mol$^{-1}$. This is natural, as it is determined predominantly by phonons and vibrations of BEDT-TTF molecules, which are very similar for the studied compounds. The difference between the two materials appears below about 6 K, where for κ-Hg-Br $C_p$ shows a linear term $\gamma = 13.8 \pm 3.1 mJK^{-2}mol^{-1}$ (inset in Fig. 3D ). Spinon excitations can produce a linear term in heat capacity [30, 31], suggesting a spin-liquid behaviour of κ-Hg-Br. For κ-Hg-Cl $\gamma$= 0 within the precision of our measurements.

An ordering of electric dipoles observed in κ-Hg-Cl does not necessarily imply magnetic order [1]. However, a theoretical proposal for a dipole order in a ``paired electron crystal'' suggest magnetic interactions as a driving force for the charge order on a frustrated dimer lattice and a spin singlet ground state [6]. A single phase transition observed at 30 K can be evidence of simultaneous electric dipole ordering and singlet formation in $\kappa$-Hg-Cl. On the other hand, the temperature of spin ordering can be lower than that of the charge order, as is observed in one-dimensional materials, and suggested by calculations [32]. Because heat capacity is found to not be always sensitive to magnetic phase transitions in these two-dimensional materials [33], further studies such as with nuclear magnetic resonance (NMR) are necessary to identify the magnetic ground state of $\kappa$-Hg-Cl. Paired electron crystal proposed as a ground state of $\kappa$-Hg-Cl [6, 10] can be regarded as a variation of a valence bond solid [6, 29]. In these terms, the quantum dipole liquid in $\kappa$-Hg-Br can be a realization of a resonant valence bond state [34].

The quantum dipole liquid was suggested as one possible explanation of the origin of the spin liquid state in $\kappa$-(BEDT-TTF)$_2$Cu$_2$(CN)$_3$ [1], however our work shows that this material does not demonstrate the signatures of this state. Its spectrum of magnetic excitations is well understood within a model of spin 1/2 on a triangular lattice with $J$=250 K (Fig. 3, Ref. [20]). There is a recent suggestion [35] that $\kappa$-(BEDT-TTF)$_2$Cu$_2$(CN)$_3$ experiences a lowering of magnetic dimensionality owing to destructive interference of magnetic interactions in one of the directions. A necessary test would be a comparison of magnetic excitation spectra of this model to available experimental Raman scattering data on magnetic excitations in $\kappa$-(BEDT-TTF)$_2$Cu$_2$(CN)$_3$ and antiferromagnetically ordered BEDT-TTF based material. At this point it is clear that the spectrum of collective excitations in $\kappa$-Hg-Br is very different from that of κ-(BEDT-TTF)$_2$Cu$_2$(CN)$_3$. Based on that we can suggest that κ-(BEDT-TTF)$_2$Cu$_2$(CN)$_3$ is a regular dimer Mott insulator with a homogeneous distribution of charge on (BEDT-TTF)$_2^{1+}$ dimer on relevant time scales. If the quantum dipole liquid model is relevant to κ-(BEDT-TTF)$_2$Cu$_2$(CN)$_3$ at all, it puts this compound quite far from a quantum phase transition into a dipole solid state. According to the model in Reference [1], tuning between a quantum dipole solid and a quantum dipole liquid can be accomplished by varying the $t_b/t_d$ ratio, where $t_b$ is an overlap integral between the dimers, and $t_d$ is an intra-dimer one. Indeed, $t_d$ is found to be larger for κ-(BEDT-TTF)$_2$Cu$_2$(CN)$_3$ than for κ-Hg-Cl [10]. The existing experimental data do not provide straightforward evidence of tuning from a dipole liquid to a dipole solid by hydrostatic or chemical pressure for κ-Hg-Br and κ-Hg-Cl family of materials. Although a charge ordered state in κ-Hg-Cl is suppressed by an external pressure of about 1 kbar [36], the unit cell of κ-Hg-Br is somewhat larger than that of κ-Hg-Cl. Calculations of the electronic structure of these materials and its change with pressure, as well as further explorations of magnetic properties are necessary for further understanding of the phase diagram.

## References

[1]  C. Hotta, *Physical Review B* **82**, 241104 (2010).

[2]  M. Naka, S. Ishihara, *Phys. Rev. B* **93**, 195114 (2016).

[3]  N. Y. Yao et al *et al.*, *Nat. Phys* (2018).

[4]  S. Rowley, *et al.*, *Nat. Phys* **10**, 367 (2014).

[5]   S.-P. Shen, *et al.*, *Nature communications* **7** (2016).

[6]   S. Dayal, R. T. Clay, H. Li, S. Mazumdar, *Phys. Rev. B* **83**, 245106 (2011).

[7]   K. Yakushi, K. Yamamoto, T. Yamamoto, Y. Saito, A. Kawamoto, *Journal of the Physical Society of Japan* **84**, 084711 (2015).

[8]   K. Sedlmeier, *et al.*, *Phys. Rev. B* **86**, 245103 (2012).

[9]   M. Aldoshina, *et al.*, *Synthetic metals* **56**, 1905 (1993).

[10]   N. Drichko, *et al.*, *Phys. Rev. B* **89**, 075133 (2014).

[11]   R. Lyubovskii, R. Lyubovskaya, O. Dyachenko, *Journal de Physique I* **6**, 1609 (1996).

[12]   T. Ivek, *et al.*, *Phys. Rev. B* **96**, 085116 (2017).

[13]   M. Dressel, , N. Drichko, *Chemical Reviews* **104**, 5689 (2004).

[14]   T. Yamamoto, *et al.*, *The Journal of Physical Chemistry B* **109**, 15226 (2005).

[15]   Y. Kim, *et al.*, *Applied Physics Letters* **100**, 071907 (2012).

[16]   K. Yakushi, *Crystals* **2**, 1291 (2012).

[17]   A. Girlando, *et al.*, *Phys. Rev. B* **89**, 174503 (2014).

[18]   T. P. Devereaux, R. Hackl, *Reviews of modern physics* **79**, 175 (2007).

[19]   N. Drichko, R. Hackl, J. A. Schlueter, *Phys. Rev. B* **92**, 161112 (2015).

[20] Y. Nakamura, et al., *Journal of the Physical Society of Japan* **83**, 074708 (2014).

[21] Y. Nakamura, T. Hiramatsu, Y. Yoshida, G. Saito, H. Kishida, *Journal of the Physical Society of Japan* **86**, 014710 (2016).

[22] Y. Shimizu, K. Miyagawa, K. Kanoda, M. Maesato, G. Saito, *Physical review letters* **91**, 107001 (2003).

[23] N. Drichko, et al., *Physica C: Superconductivity and its applications* **460**, 125 (2007).

[24] J. Merino, A. Greco, R. H. McKenzie, M. Calandra, *Phys. Rev. B* **68**, 245121 (2003).

[25] M. Dressel, N. Drichko, J. Schlueter, J. Merino, *Phys. Rev. Lett.* **90**, 167002 (2003).

[26] S. Caprara, C. Di Castro, M. Grilli, D. Suppa, *Phys. Rev. Lett.* **95**, 117004 (2005).

[27] V. Malinovsky, A. Sokolov, *Solid state communications* **57**, 757 (1986).

[28] G. Chen, L. Balents, A. P. Schnyder, *Phys. Rev. Lett.* **102**, 096406 (2009).

[29] L. Balents, *Nature* **464**, 199 (2010).

[30] S. Yamashita, et al., *Nature Physics* **4**, 459 (2008).

[31] Y. Zhou, K. Kanoda, T.-K. Ng, *Reviews of Modern Physics* **89**, 025003 (2017).

[32] M. Naka, S. Ishihara, *Scientific reports* **6** (2016).

[33] Y. Nakazawa, S. Yamashita, *Crystals* **2**, 741 (2012).

[34] P. Anderson, G. Baskaran, Z. Zou, T. Hsu, *Physical review letters* **58**, 2790 (1987).

[35] B. Powell, E. Kenny, J. Merino, *Physical Review Letters* **119**, 087204 (2017).

[36] A. Löhle, *et al.*, *Journal of Physics: Condensed Matter* **29**, 055601 (2016).

[37] H. C. Kandpal, I. Opahle, Y.-Z. Zhang, H. O. Jeschke, R. Valenti, *Phys. Rev. Lett.* **103**, 067004 (2009).

[38] M. Maksimuk, K. Yakushi, H. Taniguchi, K. Kanoda, A. Kawamoto, Journal of the Physical Society of Japan 70, 3728 (2001).

[39] F. Vernay, T. Devereaux, M. Gingras, Journal of Physics: Condensed Matter 19, 145243 (2007).

[40] N. Perkins, W. Brenig, Phys. Rev. B 77, 174412 (2008).

**Acknowledgements. Funding:** The work at the Institute of Quantum Matter was supported by the U.S. Department of Energy, Office of Basic Energy Sciences, Division of Material Sciences and Engineering under Grant No. DE-FG02-08ER46544. The work in Chernogolovka was supported by FASO Russia, state registration number 0089-2014-0036. J.A.S acknowledges support from the Independent Research and Development program from the NSF while working at the Foundation and from the National High Magnetic Field Laboratory (NHMFL) User Collaboration Grants Program (UCGP). Work at ANL was supported by University of Chicago Argonne, LLC, Operator of Argonne National Laboratory ("Argonne") Argonne, a U.S. Department of Energy Office of Science laboratory, is operated under Contract No. DE-AC02-06CH11357 **Author contributions**: N.D. conceived and designed the experiments; N.H., S.C., and N. D. performed the experiments and data analysis; J.A.S. contributed characterized samples of k-(BEDT-TTF)$_2$Cu$_2$(CN)$_3$, S.T.,E.I.Z., and R.N.L. contributed characterized samples of k-(BEDT-TTF)$_2$Hg(SCN)$_2$Br and k-(BEDT-TTF)$_2$Hg(SCN)$_2$Cl

**Competing interests:** The authors declare that they have no competing financial interests.

**Data availability:** Data presented in the paper are availavle as spread sheets in the Supplemental Information.

**Supplementary Materials:**
Experimental methods
Supplementary text
Equation S1
Figure S1-S2
Tables S1-S2

Figure 1: **Crystal structure and phases of BEDT-TTF based crystals (A)** Schematic structure of a BEDT-TTF based crystal; the molecule is highlighted in red; **(B)** Structure of a BEDT-TTF layer in the $(bc)$ plane of the $\kappa$-Hg-Cl crystal as determined from X-ray diffraction [10]. BEDT-TTF molecules are bound in dimers (circles). The dimer sites form an anisotropic triangular lattice. **(C)** Schematic structure of a BEDT-TTF layer in a dimer Mott insulator on an anisotropic triangular lattice formed by (BEDT-TTF)$_2^{1+}$ sites with S=1/2 (spins depicted by green arrows) and magnetic exchange between sites $J_M$ and $J'_M$. The model is relevant to the spin liquid candidate $\kappa$-(BEDT-TTF)$_2$Cu$_2$(CN)$_3$ with $J'_M/J_M$=0.64 [37]. **(D)** Schematic structure of a BEDT-TTF layer in case of a dipole solid (paired electron crystal [6]). Within (BEDT-TTF)$_2^{1+}$ dimer sites charge-rich and charge-poor molecules are denoted by red and blue colour, respectively. The dimer sites thus possess a dipole moment. $J_{DC}$ is the magnetic interaction between spins (marked by green arrows) on neighboring charge-rich molecules. Spins of the nearest neighbor charge-rich sites will form spin-singlets (6). This situation is relevant to $\kappa$-Hg-Cl. **(E)** Schematic structure of a BEDT-TTF layer in a quantum dipole liquid. The charge is fluctuating between the molecules in (BEDT-TTF)$_2^{1+}$ dimers, as denoted by blurry red and blue ovals, leading to electric dipoles fluctuations. Associataed spins also show fluctuations. This situation is relevant to $\kappa$-Hg-Br.

Figure 2: **Raman spectra in the region of C=C vibrations of BEDT-TTF. (A)** Temperature dependence of the κ-Hg-Cl spectra in the region of $\nu_2$ and $\nu_3$ modes. Note the splitting of $\nu_2$ mode in the dipole solid (charge ordered) state at 20 K with frequencies corresponding to BEDT-TTF $^{0.4+}$ and BEDT-TTF $^{0.6+}$. **(B)** Shape of $\nu_2$ mode calculated from the two-sites jump model, see Eq. S1. The upper spectrum is of a static system ($\omega_{ex}$=0) with bands corresponding to BEDT-TTF $^{0.4+}$ and BEDT-TTF $^{0.6+}$ as in the dipole solid state of κ-Hg-Cl. Note that on the increase of exchange frequency $\omega_{ex}$ the bands widen and move close to each other. The lower two spectra at $\omega_{ex}$= 30 and 40 cm$^{-1}$ reproduce the $\nu_2$ shape of κ-Hg-Br at 8 and 35 K correspondingly, taking into account the natural width $\Gamma$ for the relevant temperature. **(C)** Temperature dependence of the κ-Hg-Br spectra in the region of $\nu_2$ and $\nu_3$ modes. The $\nu_2$ band does not split but shows some widening at lowest temperature. **(D)** Temperature dependence of centerfrequency (upper panel) and line width (lower panel) for $\nu_2$ (triangles) and $\nu_3$ (diamonds) modes for κ-Hg-Br. The line width of $\nu_2$ for κ-Hg-Br goes through a minimum at around 80 K, whereas that of $\nu_3$ decreases continuously. **(E)** Illustration of the movements of atoms associated with $\nu_2$ and $\nu_3$-mode vibrations in a BEDT-TTF molecule.

Figure 3: **Temperature dependence of Raman spectra.** Shown are the spectra for **(A)**

κ-Hg-Br and **(B)** κ-Hg-Cl in $A_{1g}$ symmetry in the frequency range between 0 and 300 cm$^{-1}$. Phonons are found at similar frequencies for both compounds. In the spectra of κ-Hg-Br a background develops at temperatures below 100 K. Spectra at 300 and 11 K for κ-Hg-Br for $B_{1g}$ symmetry are shown in the inset to (A). In $B_{1g}$ scattering channel the low frequency background, interpreted later as a collective mode (see text), shows strong coupling to the phonons. **(C)** Temperature dependence of the collective mode in $A_{1g}$ scattering channel for κ-Hg-Br determined by substracting phonons from the full Raman spectrum, see SI for the details of the precedure. Inset : temperature dependence of the normalized intensity of the collective mode. **(D)** Temperature dependence of the heat capacity $C_p$ for κ-Hg-Cl (red line) and κ-Hg-Br (black line) below 40 K. The two curves deviate from each other below approximately 6 K. Inset: Low temperature data with linear behaviour of heat capacity for κ-Hg-Br . **(E)** Raman spectra in $B_{1g}$ polarization at 20 K in the range between 800 and 1100 cm$^{-1}$ with phonons and luminescence background subtracted for κ-(BEDT-TTF)$_2$Cu$_2$(CN)$_3$ (upper panel) and κ-Hg-Cl and κ-Hg-Br (lower panel). See details on the subrtaction procedure in SI. Schematic pictures of the relevant models with different charge distibution are shown. The spectra of the dimer Mott insulator on the triangular lattice κ-(BEDT-TTF)$_2$Cu$_2$(CN)$_3$ (upper panel) demonstrate magnetic excitations below approx. 600 cm$^{-1}$ . This feature is absent in the spectra of both κ-Hg-Br (black) and κ-Hg-Cl (red). The increase of intensity in the spectra of κ-Hg-Br below 200 cm$^{-1}$ is caused by the collective mode fully shown in (A).

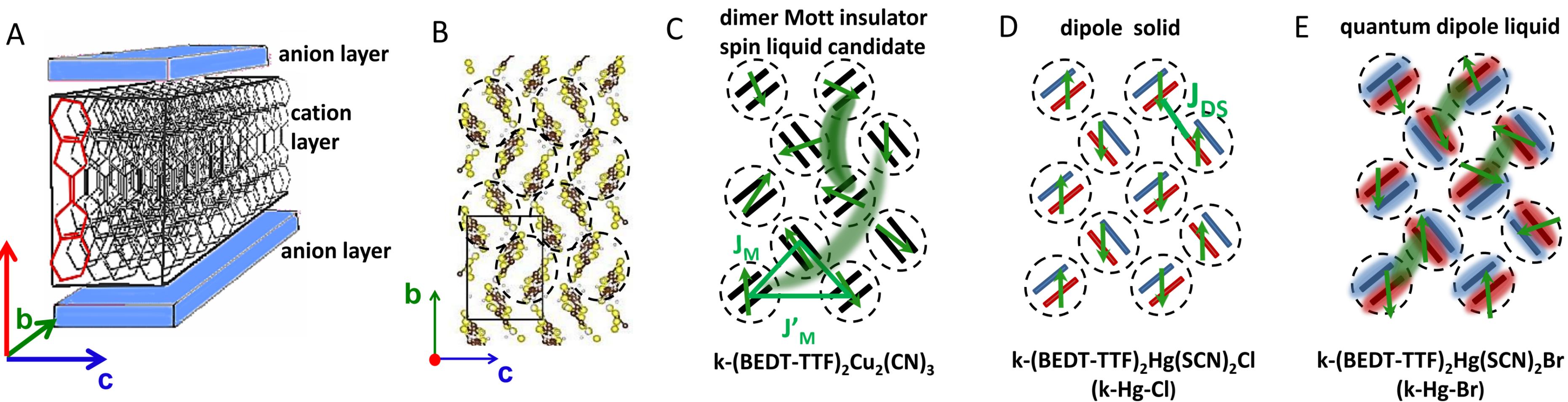

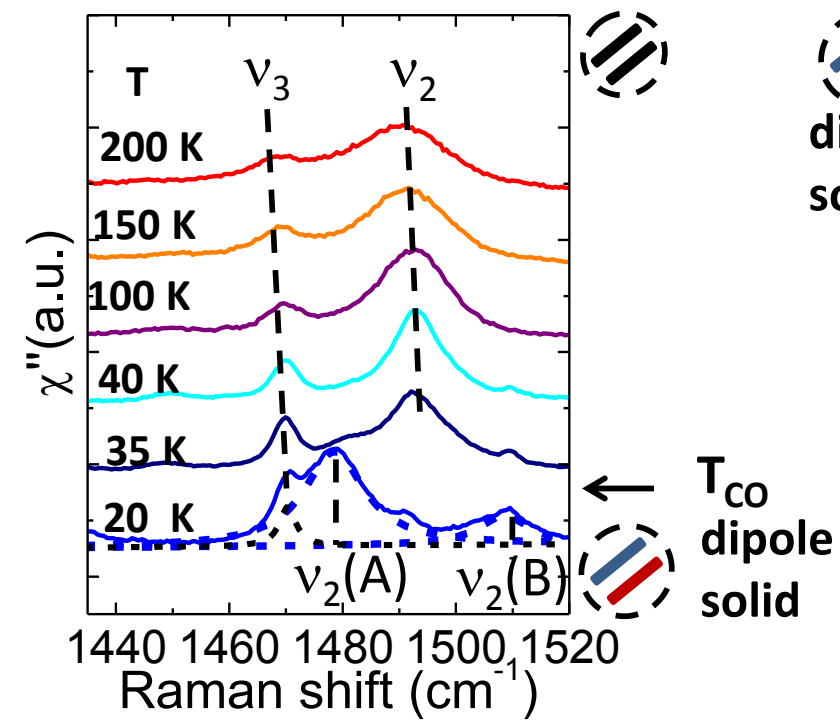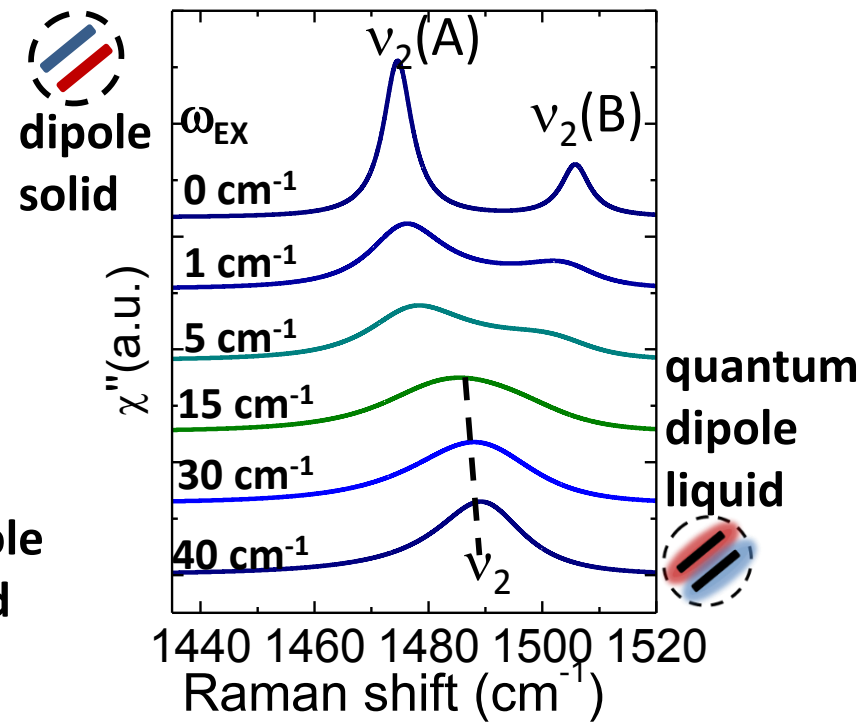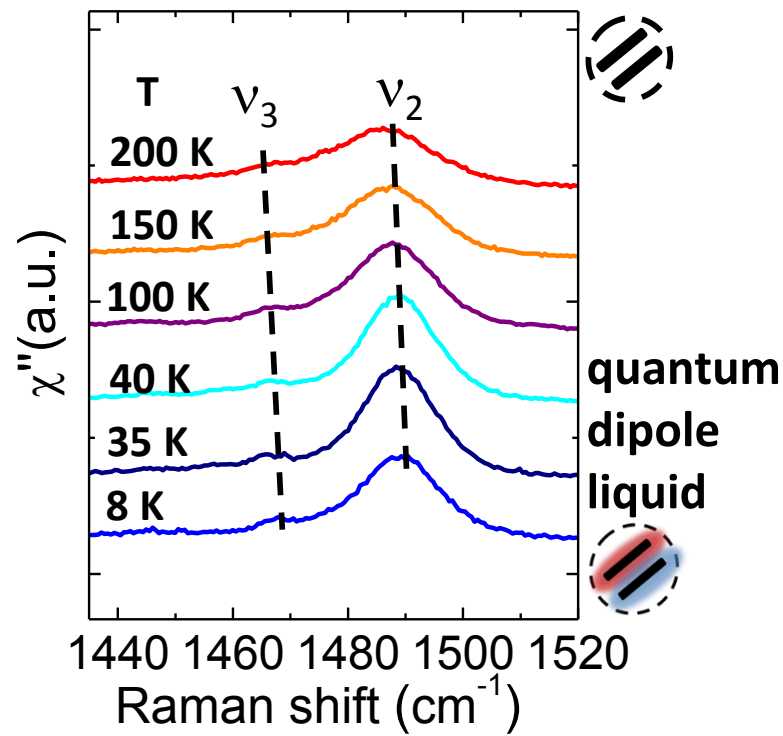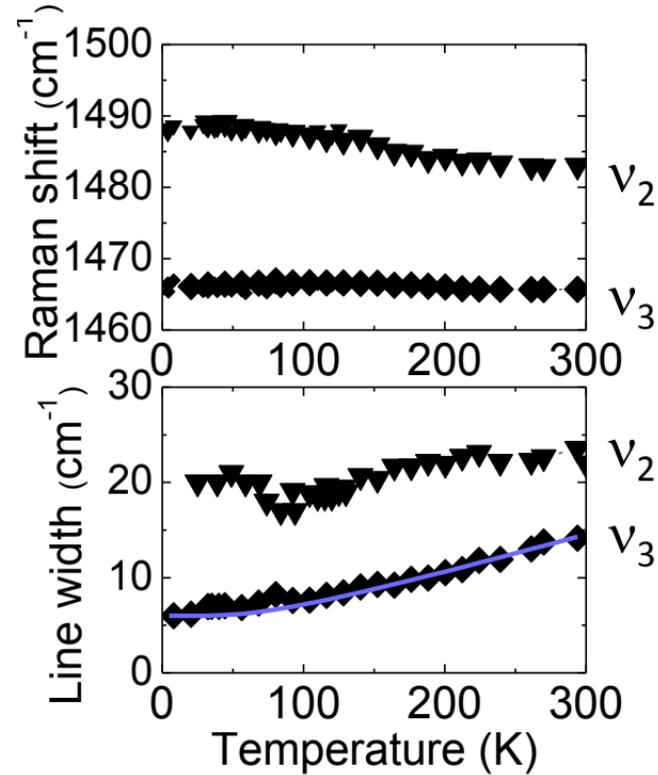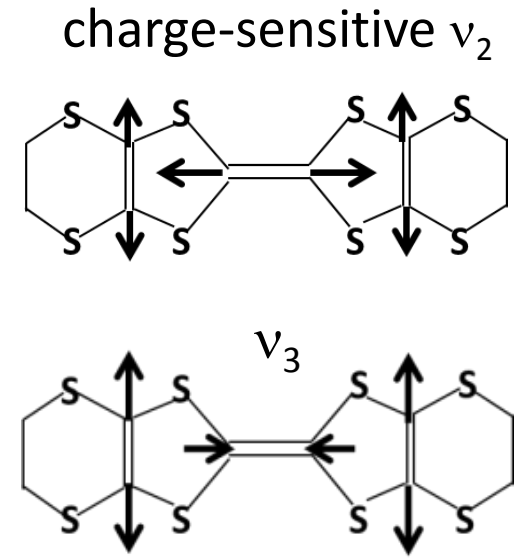

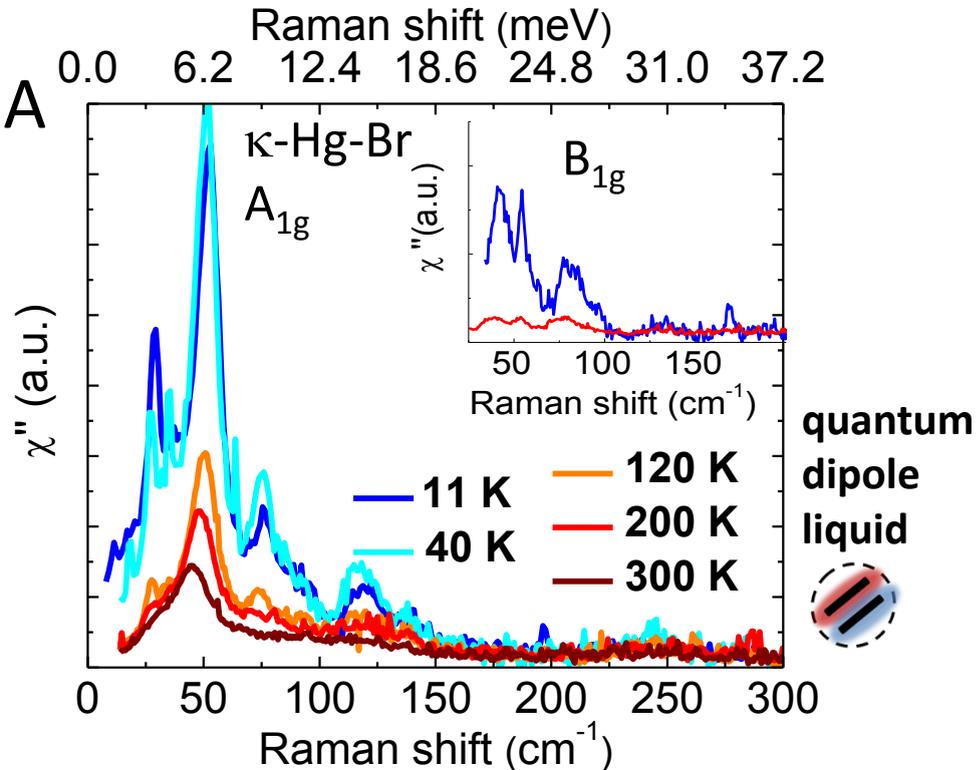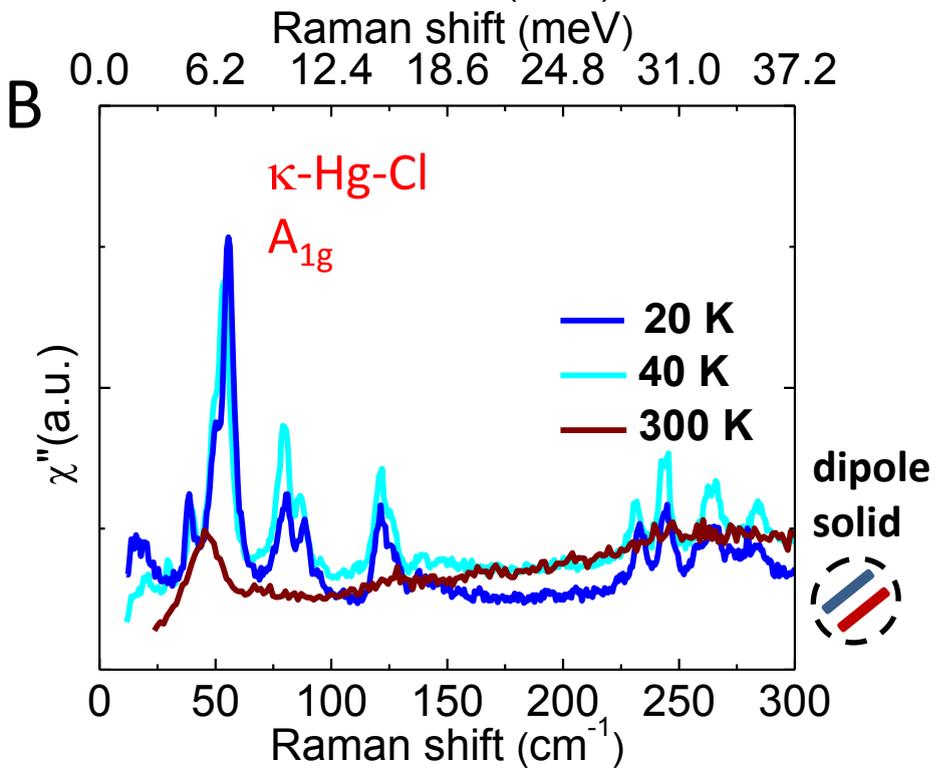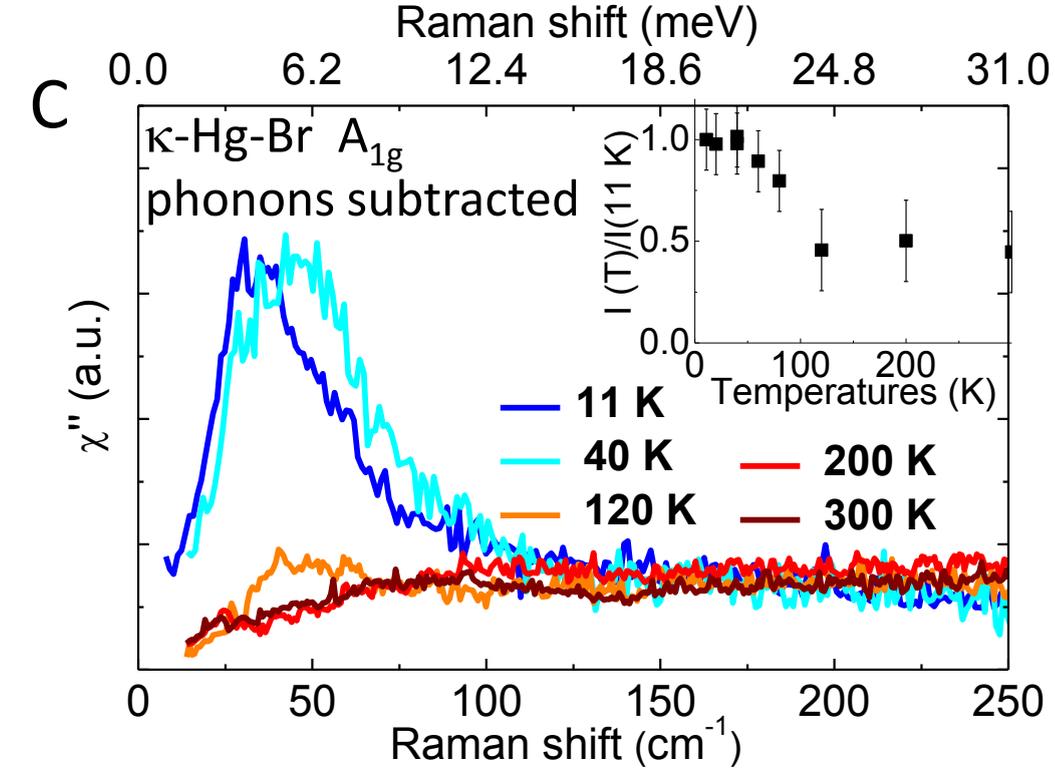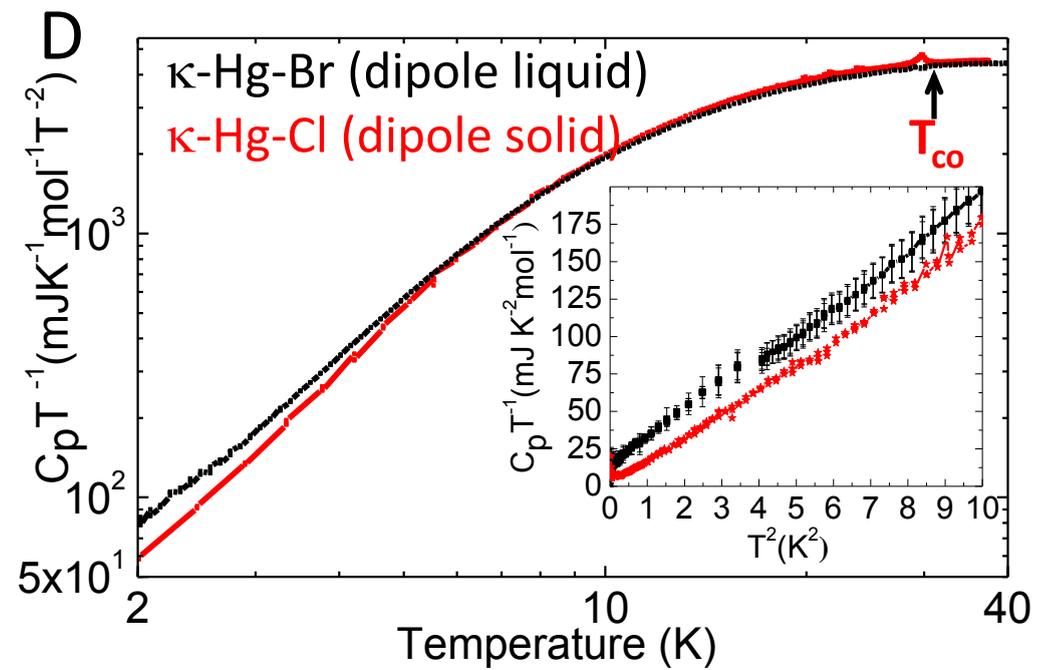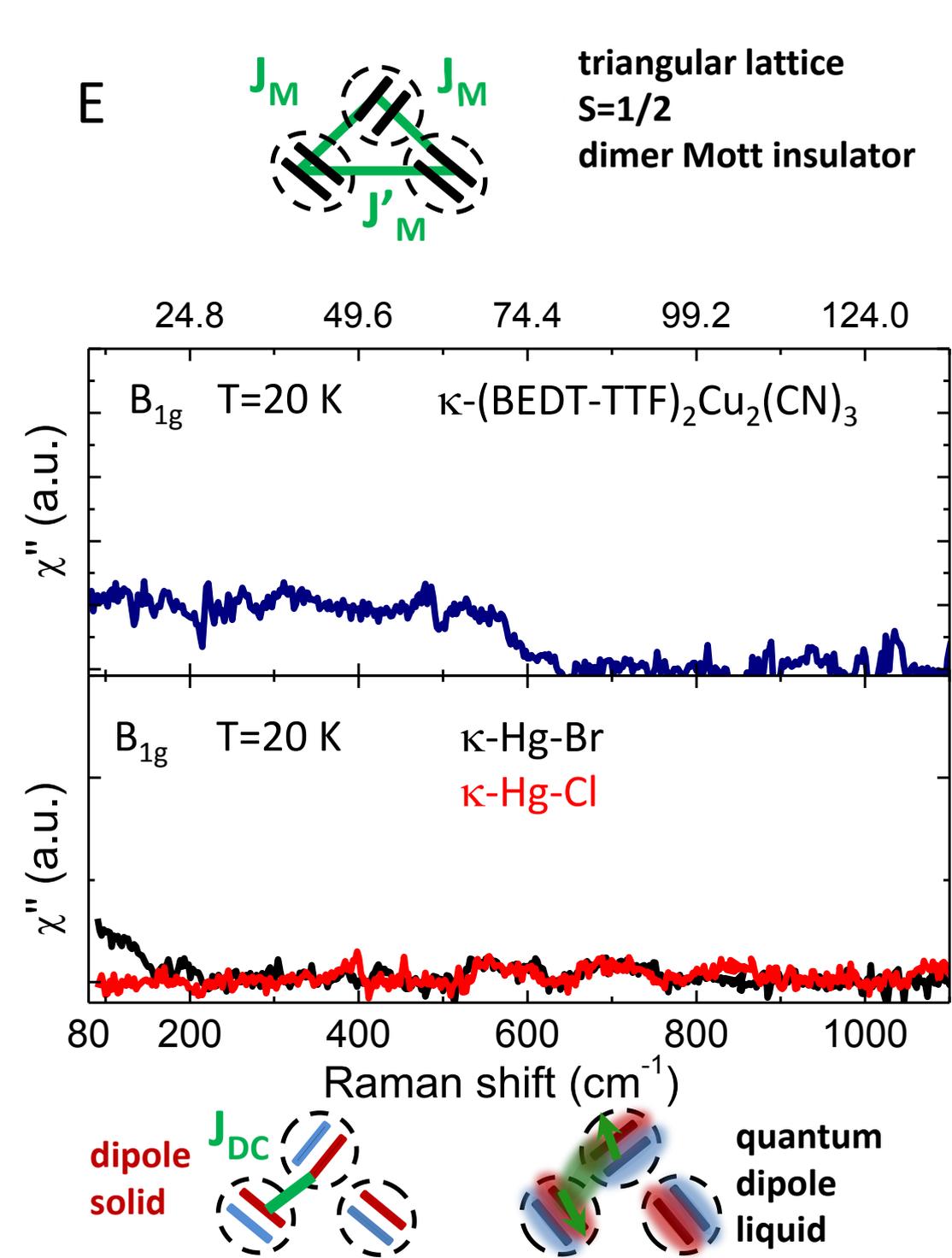

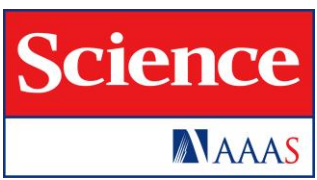

# Supplementary Materials for

## Evidence for a quantum dipole liquid state in an organic quasi-two-dimensional material

Nora Hassan, [1] Streit Cunningham, [1] Martin Mourigal, [2] Elena I. Zhilyaeva, [3] Svetlana A. Torunova, [3] Rimma N. Lyubovskaya, [3] John Schlueter, [4,5] Natalia Drichko [1*]

[1] The Institute for Quantum Matter and the Department of Physics and Astronomy, The Johns Hopkins University, Baltimore, Maryland 21218, USA
[2] School of Physics, Georgia Institute of Technology, Atlanta, GA 30332, USA
[3] Institute of Problems of Chemical Physics, Chernogolovka, Russia
[4] Division of Materials Research, National Science Foundation, Alexandria, VA 22314, USA
[5] Materials Science Division, Argonne National Laboratory, Argonne, IL 60439, USA

correspondence to: drichko@jhu.edu

**This PDF file includes:**

Materials and Methods
Supplementary Text
Figs. S1 to S2
Tables S1 to S2

**Other Supplementary Materials for this manuscript includes the following:**

Databases S1 to Sx as zipped archives:
kHgCl_Fig_2A.pdf
kHgBr_Fig_2B.pdf
kHgBr_Fig_2C.pdf
kHgBr_Fig_3A.pdf
kHgBr_Fig_3B.pdf
kHgBr_Fig_3C.pdf
Fig3_E.pdf

**Materials and Methods**

Synthesis

Single crystals of κ-(BEDT-TTF)$_2$Hg(SCN)$_2$Cl (κ-Hg-Cl) and κ-(BEDT-TTF)$_2$Hg(SCN)$_2$Br (κ-Hg-Br) were prepared by electrochemical oxidation of the BEDT-TTF solution in 1,1,2-trichloroethane (TCE) at a temperature of 40$^o$ C and a constant current of 0.5 µA. A solution of Hg(SCN)$_2$, [Me4N]SCN · KCl, and dibenzo-18-crown-6 in 1:0.7:1 molar ratio in ethanol/TCE was used as supporting electrolyte for the κ-Hg-Cl preparation. For the κ-Hg-Br preparation, a supporting electrolyte Hg(SCN)$_2$/[Me4N]SCN · 1.5KBr/ dibenzo-18-crown-6 in 1:0.4:1 molar ratio was used. The composition of the crystal was verified by electron probe microanalysis and X-ray diffraction.

Raman scattering measurements

Raman scattering was measured in pseudo-Brewster angle geometry using T64000 triple monochromator spectrometer equipped with the liquid N$_2$ cooled CCD detector. For the measurements in the 5-400 cm$^{-1}$ range T64000 in triple monochromator configuration was used, for the measurements in the range 80-2000 cm$^{-1}$ single monochromonator configuration with an edge filter option was used. Spectral resolution was 2 cm$^{-1}$. Lines of Ar$^+$-Kr$^+$ Coherent laser at 514 nm and 647 nm were used for excitation. Laser power was kept at 2 mW for the laser probe size of approximately 50 by 100 µm. This ensured that laser heating of the sample was kept below 2 K, as was proved by observing the temperature of ordering transition in κ-Hg-Cl. Measurements at temperatures down to 5 K were performed using Janis ST500 cold finger cryostat. Samples were cooled from 300 K with cooling rates between 0.2 and 0.5 K/min. Samples were glued on the cold finger of the cryostat using GE varnish. The experiments were performed on at least 6 samples to ensure reproducibility of the results. A few wavenumbers spread in the width of $\nu_2$ band at low temperatures for κ-Hg-Br was associated with a weak strain originating from GE varnish, all the other parameters of the spectra were reproducible within the error bar of the measurements. The crystals were oriented using polarization-dependent Raman scattering measurements. For the measurements, electrical vector of excitation $e_L$ and scattered $e_S$ light were polarized along b and c axes. Our notations of polarizations refer to the structure and symmetry of the BEDT-TTF layer, to make an easy comparison to the calculations which refer to D$_{4h}$ *(18)* without losing the information about the symmetry of the real crystal. Thus A$_{1g}$ symmetry corresponds to the measurement in *(b, b)* and *(c, c)* geometries, and B$_{1g}$ corresponds to *(b, c)* and *(c, b)* geometries. All spectra were corrected by the Bose-Einstein thermal factor.

Raman scattering data analysis

Intensity of the collective mode was calculated as

$$I(T) = \int_0^{200 cm^{-1}} \chi''(\omega,T) d\omega$$

The calculations of the shape of the $\nu_2$ band depending on $\omega_{EX}$ jump rate between sites are done using Eq. S1 *(16)*.

$$I(\omega) \propto \mathrm{Re}[(\alpha_R, \alpha_P)\begin{pmatrix} i(\omega-(\omega_{1/2}-\Delta))+\omega_{EX}/2+\Gamma/2 & -\omega_{EX}/2 \\ -\omega_{EX}/2 & i(\omega-(\omega_{1/2}+\Delta))+\omega_{EX}/2+\Gamma/2 \end{pmatrix}^{-1}\begin{pmatrix}\alpha_R \\ \alpha_P\end{pmatrix}]$$

(1)

Here $\Delta = \dfrac{\nu_2(\mathrm{BEDT\text{-}TTF}^{+0.4}) - \nu_2(\mathrm{BEDT\text{-}TTF}^{+0.6})}{2} = 16$ cm$^{-1}$ is obtained using frequencies of $\nu_2$ of these two species observed in the ordered phase of κ-Hg-Cl. $\Gamma$ is the natural width of the vibrational bands experimentally demonstrated in our measurements by the width of $\nu_3$. $\omega_{EX}$ is the frequency of the jumps between the two states, BEDT-TTF$^{0.4+}$ and BEDT-TTF$^{0.6+}$. $\alpha_R$ and $\alpha_P$ are intensities of $\nu_2$ (BEDT-TTF$^{0.4+}$) and $\nu_2$ (BEDT-TTF$^{0.6+}$) obtained from the spectra of κ-Hg-Cl in the ordered state. List of parameters which correspond to the calculated spectra presented in Fig. 2 of the paper are shown in Table S1.

Raman scattering spectra of all the discussed compounds show an overlap of electronic and magnetic components, phonons, and lattice vibrations *(13)*, and luminescence which appears at frequencies above 3000 cm$^{-1}$ *(10)*. Basing on the fits of these different known contributions, their dependence on temperature, polarization, and chemical composition of the samples, we manage to separate these contributions. In Fig. S1 we demonstrate the different contributions to the spectra and the results of the subtraction.

Fig. S1 A shows the spectral region of the collective mode with the raw data for κ-Hg-Br at T=11 K. In the middle panel we compare phonon contribution for κ-Hg-Br (back line) to the spectra of κ-Hg-Cl (red line). As expected for iso-structural compounds, phonon spectra of these materials are very similar, with the differences introduced by coupling of phonons of κ-Hg-Br to the collective mode. The lower panel shows the collective mode as it is presented in the paper.

Fig. S1 B shows the spectra of κ-Hg-Br and κ-Hg-Cl in the wide range where magnetic excitations of κ-(BEDT-TTF)$_2$Cu$_2$(CN)$_3$ appear. Upper panel shows raw data, and luminescence and electronic contributions, which are received from the fitting of the spectra. A restriction on the fit parameters comes from comparing data at the different polarizations and temperatures. Middle panels show phonon contribution. For κ-Hg-Cl and κ-Hg-Br those are nearly equal, the differences appear due to electron-phonon coupling. Also, some differences in electronic contribution appear due to different ground states. An asymmetric feature in κ-Hg-Cl spectra at around 400 cm$^{-1}$ is due to and electronic gap in charge order state or electron-phonon coupling. The lower panel shows the resulting spectra presented in the paper.

Fig. S1 C Upper panel shows both A$_{1g}$ and B$_{1g}$ scattering channels κ-(BEDT-TTF)$_2$Cu$_2$(CN)$_3$ together with the luminescence (blue dotted line) and electronic contribution (red line) with maximum at around 1500 cm$^{-1}$. While separation of the contributions is straightforward in B$_{1g}$, the intense electronic band in A$_{1g}$ channel introduces a large error bar in the estimate for magnetic excitations. This is the reason why their intensity was found very low, for example in Ref. *(20)*. We show the error bar in the lower panel where magnetic excitations for the both polarization are demonstrated.

Heat capacity measurements

Heat capacity was measured using Quantum Design PPMS system equipped with the DR option. All data were collected on warming, after cooling with the cooling rate of 0.5 K/min.

Crystal structure of the samples

Crystal structures of the studied materials κ-Hg-Br and κ-Hg-Cl were published in Ref. *(9)* and *(10)*. In the Table S2 we present basic information about both structures. It shows that the volume of the unit cell for κ-Hg-Br compound is somewhat larger than κ-Hg-Cl. Calculations of the electronic structure are necessary to understand how exactly this change of geometry between two compounds will affect parameters which define the ground state.

**Supplementary Text**

Temperature dependence of vibrational modes for κ-Hg-Cl

Both $\nu_2$ and $\nu_3$ have large intensity in $A_{1g}$ and $B_{1g}$ symmetries. In Fig. S2 we present temperature dependence of parameters of $\nu_3$ and charge-sensitive $\nu_2$ vibrations for κ-Hg-Cl. $\nu_3$ band shows normal hardening on cooling and a decrease of the width, with the temperature dependence coinciding with that for κ-Hg-Br. If no coupling to electronic or magnetic excitations is present, the main process which defines relaxation of a phonon excited state is scattering on lower frequency phonons with frequency denoted here $\omega_0$. Respectively, a decrease of a line width of a phonon on cooling is determined by a thermal population of the phonon levels $\omega_0$ involved in the scattering process:

$\Gamma(\omega) = \Gamma_0 + \alpha (e^{\frac{\hbar \omega_0}{k_B T}} - 1)^{-1}$, where $\Gamma_0$ is temperature-independent scattering rate defined by disorder, and $\alpha$ is a probability of the decay *(15, 38)*. A decrease of the width of $\nu_3$ is described well by this formula with $\omega_0 = 150$ cm$^{-1}$, which suggest the most probable decay processes. Our results coincide with that for κ-(BEDT-TTF)$_2$Cu[N(CN)$_2$]Br in Ref. *(38)*.

The width of the charge-sensitive $\nu_2$ in the spectra of κ-Hg-Cl decreases below that of κ-Hg-Br on cooling down to approximately 40 K, and then shows a weak increase till the compound reaches the ordering transition at 30 K, and decreases again below the transition as expected in the ordered state.

Temperature dependence of vibrational modes for κ-Hg-Br

In the spectra of κ-Hg-Br $\nu_2$ mode hardens on cooling similar to κ-Hg-Cl. At temperatures between 150 and 100 K we observe a small change of a slope of the temperature dependence of the position of the $\nu_2$ mode. This change occurs still in the metallic state *(12)*. The change is close to the size of an error with which we determine parameters of the bands. Here the size of triangle/rhombs with which the data are plotted depict the size of the error bars. A similar change of a hardening slope was also noted, for example, for κ-(BEDT-TTF)$_2$Cu[N(CN)$_2$]Br in Ref. *(38)*. A reason for the change of the slope in the temperature dependence of the band frequency could be the following (i) As mentioned in Ref. *(38)*, hardening depends on the compression of the lattice, thus changes in the latter can define it. At this point no such precise temperature dependence of the crystal structure has been measured. (ii) While there is a good understanding of the dependence of the frequency of $\nu_2$ mode on the charge distribution on the lattice in the insulating state, at this point it is not clear how this frequency would change due to interactions of this phonon with itinerant electrons. These interactions can be a reason for the change of behavior.

We used Eq. S1 to reproduce $\nu_2$ line shapes for low temperature κ-Hg-Cl and κ-Hg-Br and also some line shapes for possible exchange rates which are not present in experimental data to illustrate the evolution of the line shape in the "two-sites jump model". The results of our calculations are presented in Fig. 2B of the main text. Parameters for which the calculations are done are listed in the Table S1.

Raman scattering data: Low-frequency spectra of κ-Hg-Br and κ-Hg-Cl in $B_{1g}$ symmetry

In Fig. S2 B we present temperature dependence of the Raman scattering spectra for κ-Hg-Br and κ-Hg-Cl compounds for $B_{1g}$ symmetry in the spectral region between 25 and 400 cm$^{-1}$. It shows that (i) Similar to $A_{1g}$ for κ-Hg-Br, the background feature due to the collective mode at about 40 cm$^{-1}$ appears also in spectra of κ-Hg-Br in $B_{1g}$ symmetry channel. (ii) The low frequency phonons in κ-Hg-Br $B_{1g}$ spectra remain broad, which can be explained by an interaction of these phonons with the collective mode. This is in contrast with spectra of κ-Hg-Cl and κ-Hg-Br in $A_{1g}$ polarization, where at low frequencies narrow phonon modes are superimposed on the collective mode background. (iii) Some drop of intensity in κ-Hg-Cl spectra between 35 and 20 K is due to an opening of a gap due to an ordering transition.

Polarization dependence is an important way to characterize excitations observed in Raman scattering *(18)*. For $D_{4h}$ symmetry a clear separation of $A_{1g}$ *($x^2 + y^2$)*, $B_{1g}$ *($x^2 - y^2$)*, and $B_{2g}$ *(xy)* polarization is expected for electronic and magnetic excitations. Spectra of some materials can be mapped on $D_{4h}$ symmetry and understood within it even if the symmetry of the unit cell is lower. A good example is the antiferromagnetically ordered organic compound κ-(BEDT-TTF)$_2$Cu[N(CN)$_2$]Cl, where in a monoclinic unit cell (BEDT-TTF)$_2^{1+}$ dimers form a weakly frustrated square lattice, and two-magnon Raman excitations in that material follow the symmetry selection rules expected for $D_{4h}$. In the materials studied in this work, the symmetry of the unit cell is monoclinic as well, however it was shown *(10)* that the (BEDT-TTF)$_2^{1+}$ dimers in *(bc)* plane form slightly anisotropic triangular lattice. For $D_{3h}$ point group Raman-active symmetries cannot be fully separated by selecting polarizations. It was shown by numerical calculations, too, that magnetic excitations on a triangular lattice lose their anisotropy between $A_{1g}$ and $B_{1g}$ *(39, 40)*, a similar change is expected for electronic excitations. This explains the presence of the collective mode in spectra of κ-Hg-Br in $A_{1g}$, $B_{1g}$ *(xy),* and *($x^2 - y^2$),* which would exist as $B_{2g}$ only in $D_{4h}$. On the other hand, polarization dependence of phonons follows the full symmetry of the lattice. Our experimental data suggest that the coupling to the collective mode is larger for $B_{1g}$ phonons than for $A_{1g}$. This fact can be a realization of the expected anisotropy of the electronic excitation.

Raman scattering data: Magnetic excitations in κ-(BEDT-TTF)$_2$Cu$_2$(CN)$_3$.

In the main text of the paper we present the data on the magnetic excitations we observe in the spectra of κ-(BEDT-TTF)$_2$Cu$_2$(CN)$_3$ at low temperatures in $B_{1g}$ symmetry below 600 cm$^{-1}$. For triangular lattice materials, according to the calculations, the anisotropy of the magnetic response between $B_{1g}$ and $A_{1g}$ channel should be reduced, and eventually disappear for a case of isotropic triangular lattice *(39,40)*.

The question of the anisotropy of the magnetic response for κ-(BEDT-TTF)$_2$Cu$_2$(CN)$_3$ is not straightforward, as we discussed in the Experimental Methods section. Due to the large intensity of an electronic contribution, the origin of which still has to be understood, the error bar for the intensity of the magnetic contribution in A$_{1g}$ channel is very large. In Fig. S1 C lower panel we present a comparison of both A$_{1g}$ and B$_{1g}$ contributions after applying the described procedure. We observe a non-negligible magnetic contribution in A$_{1g}$ polarization. Detailed measurements of polarization dependence can provide more data to reduce the large error bar for the intensity of magnetic excitations in A$_{1g}$.

Heat capacity

In Fig. S2 C we compare our results on temperature dependence of heat capacity for κ-Hg-Br and κ-Hg-Cl with that received by us for κ-(BEDT-TTF)$_2$Cu$_2$(CN)$_3$. This plot demonstrates the high quality of the data obtained by us, where the κ-(BEDT-TTF)$_2$Cu$_2$(CN)$_3$ data coincide with the literature. Within the precision of the measurements we performed, of κ-(BEDT-TTF)$_2$Cu$_2$(CN)$_3$ is similar to the published. Heat capacity temperature dependence is very similar to that of κ -Hg-Br, which can be explained by a similar structure and thus phonon DOS of the compounds *(30)*.

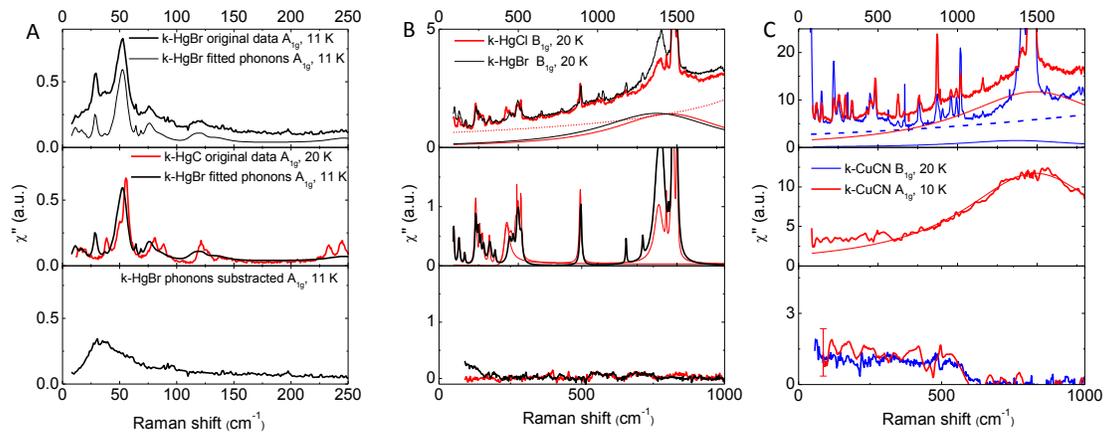

**Fig. S1.**
Illustration of a procedure of subtracting of phonon and luminescence contributions from the raw data. **(A)** Spectra of κ-Hg-Br in the range of the collective mode (0 -250 cm$^{-1}$). Raw data and phonon contribution are shown in the upper panel. Middle panel shows a comparison of phonon contribution for κ-Hg-Br to the spectra of κ-Hg-Cl in this range. Lower panel shows the collective mode contribution. **(B)** Upper panel shows raw data for κ-Hg-Br (black line) and κ-Hg-Cl (red line) at 20 K, as well as luminescence contribution, similar for both, and electronic contribution. Middle panel compared phonon contribution for κ-Hg-Br and κ-Hg-Cl. Lower panel show spectra with all these contributions subtracted. **(C)** Upper panel shows raw data for κ-(BEDT-TTF)$_2$Cu$_2$(CN)$_3$, as well as luminescence and electronic contributions. Middle panel shows the phonon contribution, and lower panel shows spectrum of magnetic excitations, revealed when all the mentioned contributions are subtracted from the raw data.

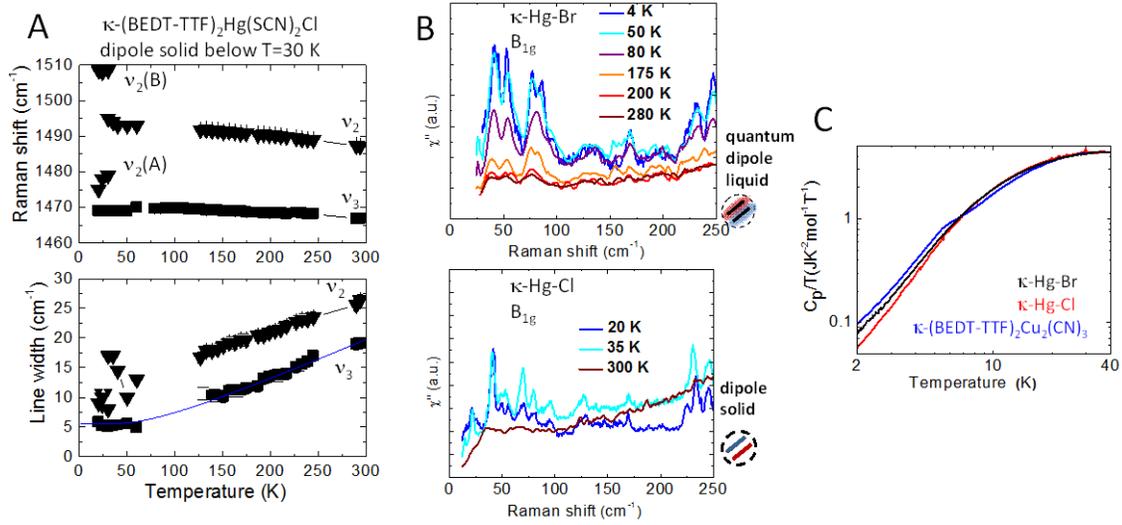

**Fig. S2**

Additional data on temperature dependence of heat capacity and Raman scattering of κ-Hg-Br and κ-Hg-Cl. **(A)** Temperature dependence of parameters of $\nu_2$ and $\nu_3$ vibrational modes for κ-Hg-Cl. Blue line in the lower panel shows a fit of the width of $\nu_3$ by

$\Gamma(\omega) = \Gamma_0 + \alpha(e^{\frac{\hbar\omega_0}{k_B T}} - 1)^{-1}$ with $\Gamma_0$=5.5 cm$^{-1}$, α=15, $\omega_0$=150 cm$^{-1}$. **(B)** Temperature dependence of $B_{1g}$ *(bc)* spectra for κ-Hg-Br (upper panel) and κ-Hg-Cl (lower panel). Note a collective mode which appears at low frequencies in the spectra of κ-Hg-Br, similar to $A_{1g}$ *(b,b)* polarization. The phonons are widened by the interaction with the collective mode. **(C)** Comparison of temperature dependence of heat capacity of κ-Hg-Br and κ-Hg-Cl with that of κ-(BEDT-TTF)$_2$Cu$_2$(CN)$_3$.

**Table S1.**

List of the parameters used in the formula Eq. S1 for the calculations of the line shapes presented in Fig. 2B of the paper

| ω, cm$^{-1}$ | ω, cm$^{-1}$ | Δ, cm$^{-1}$ | Γ, cm$^{-1}$ | α$_P$ | α$_R$ |
|---|---|---|---|---|---|
| 0 | 1490 | 15.5 | 6 | 2.05 | 3.56 |
| 1 | 1490 | 15.5 | 6 | 2.05 | 3.56 |
| 5 | 1490 | 15.5 | 6 | 2.05 | 3.56 |
| 15 | 1490 | 15.5 | 6 | 2.05 | 3.56 |
| 30 | 1490 | 15.5 | 6 | 2.05 | 3.56 |
| 40 | 1490 | 15.5 | 7 | 2.05 | 3.56 |

Table S2.
Crystal structure data for κ-Hg-Br and κ-Hg-Cl

| Formula | κ-(BEDT-TTF)$_2$Hg(SCN)$_2$Br (κ-Hg-Br) *(9)* | κ -( BEDT-TTF)$_2$Hg(SCN)$_2$Cl (κ-Hg-Cl) *(10)* |
|---|---|---|
| Space group | C2/c | C2/c |
| a (Å) | 37.09(1) | 36.9564 |
| b (Å) | 8.338(3) | 8.2887(2) |
| c (Å) | 11.738(5) | 11.7503(3) |
| α (deg) | 90 | 90 |
| β (deg) | 89.71 | 90.067 |
| γ (deg) | 90 | 90 |
| V (Å$^3$) | 3570.2(8) | 3564.29 |
| Z | 4 | 4 |